# Schwarzites for Natural Gas Storage: A Grand-Canonical Monte Carlo Study

Daiane Damasceno Borges[1], Douglas S. Galvao[1]

[1]*Applied Physics Department and Center of Computational Engineering and Science, University of Campinas - UNICAMP, Campinas-SP 13083-959, Brazil.*

ABSTRACT

*The 3D porous carbon-based structures called Schwarzites have been recently a subject of renewed interest due to the possibility of being synthesized in the near future. These structures exhibit negatively curvature topologies with tuneable porous sizes and shapes, which make them natural candidates for applications such as $CO_2$ capture, gas storage and separation. Nevertheless, the adsorption properties of these materials have not been fully investigated. Following this motivation, we have carried out Grand-Canonical Monte Carlo simulations to study the adsorption of small molecules such as $CO_2$, CO, $CH_4$, $N_2$ and $H_2$, in a series of Schwarzites structures. Here, we present our preliminary results on natural gas adsorptive capacity in association with analyses of the guest-host interaction strengths. Our results show that Schwarzites P7par, P8bal and IWPg are the most promising structures with very high $CO_2$ and $CH_4$ adsorption capacity and low saturation pressure (<1bar) at ambient temperature. The P688 is interesting for $H_2$ storage due to its exceptional high $H_2$ adsorption enthalpy value of -19kJ/mol.*

## INTRODUCTION

Schwarzites are 3D porous carbon-based structures first idealized by Mackay and Terrones [1]. These crystalline structures consist of $sp^2$ carbon atoms forming hexa-, hepta- and octagons with a negative Gaussian curvature, similar to those found in triply periodic minimal surfaces. Although these structures have not been synthesized yet, recent experimental work with graphene foams revealed similar features to Schwarzites [2], opening a possibility of their syntheis in a near future. From the theoretical point of view, these structures have been subject of many studies due to their interesting mechanical [3] and electronic [4] properties. These materials also present high porosity and elevate surface area, which make them natural candidates for adsorption applications such as $CO_2$ capture, natural gas storage and separation. In this context, they have the advantage of being very hydrophobic, which could be interesting for removing natural gas from humidified environments, such as in pos- and pre-combustion process.

Many efforts are still been devoted to search for nanoporous materials specifically targeted for natural gas storage. In particular, $H_2$ and $CO_2$ storages are the most investigated due to their relevance in $H_2$ fuel technology development and carbon dioxide emission reduction. One of the most exploited routes consists of engineering materials with a large surface area and with a strong gas affinity with the aim of enhancing the storage capacity. In this work, we propose to computationally investigate the adsorption properties of a series of Schwarzites materials for gas storage applications. We have examined four families of Schwarzites: the Primitive (P), Diamond (D), Gyroid (G) and I-Wrapped Package graph (IWPg) surfaces (see Figure 1). We have applied classical force field- Grand Canonical Monte Carlo (GCMC) simulations to predict the adsorption isotherms of $CO_2$, $CH_4$, CO, $N_2$ and $H_2$ in these solids, as well as to obtain information on the energetics of the host/guest interactions. The relation between adsorption performance and pore topology is also discussed.

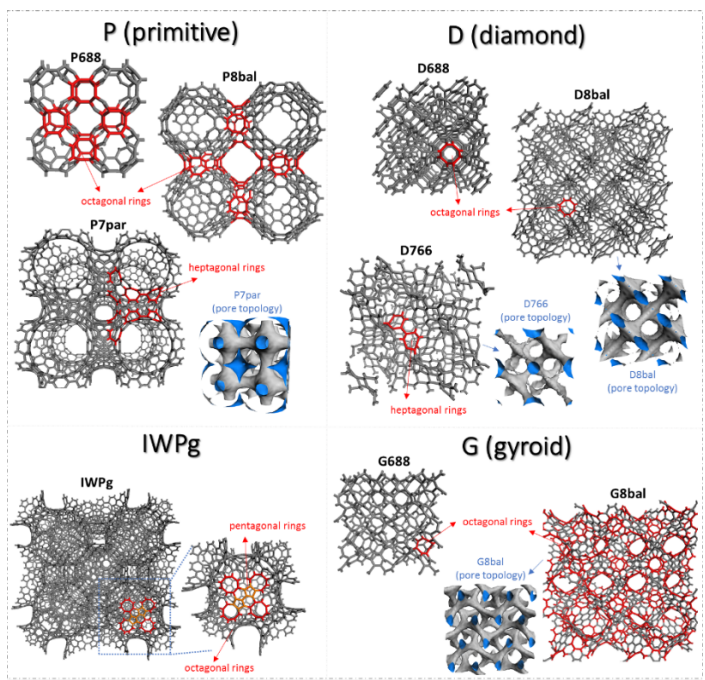

Figure 1: The Schwarzites families considered in this work: Primitive (P), Diamond (D), Gyroid (G) and I-Wrapped Package graph (IWPg) surfaces.

**METHODS**

The atomic coordinates for each framework were extracted from H. Terrones and M. Terrones publication [5]. Grand Canonical Monte Carlo (GCMC) simulations were further performed to predict the adsorption of single components, $CO_2$, $N_2$, $CH_4$, $H_2$ and CO inside each structure. The interaction between the framework and the guest species was modelled using the 12-6 Lennard-Jones (LJ) taken from the Universal force field (UFF) [6]. The $CO_2$ molecule was represented by the conventional rigid linear triatomic model, with three charged LJ interaction sites (C−O bond length of 1.149 Å) located on each atom as previously derived by Harris and Yung [7], The $N_2$ and CO molecules was also described by a three charged sites model taken from the TraPPE forcefield [8] and from the paper of Straub *et al.* [9], respectively. The $H_2$ molecules were modelled with uncharged two-sites LJ [10] and $CH_4$ was described by the TraPPE uncharged single LJ interacting site model [11]. The LJ crossing parameters for guest/MOFs interactions were obtained using Lorentz−Berthelot mixing rules. The Ewald summation was used for the calculations of the electrostatic interactions while the short-range contributions were computed with a cutoff distance of 14 Å. Gas-phase fugacity values were calculated with the Peng-Robinson equation of state [12]. These GCMC simulations were performed using CADSS (Complex Adsorption and Diffusion Simulation Suite) [13]. For each state point, $2\times10^7$ Monte Carlo steps were used for both equilibration and production runs and the adsorption enthalpy at low coverage for each gas was calculated through configurational-bias Monte Carlo simulations performed in the μVT ensemble using the revised Widom's test particle insertion method [14].

The following structural analyses were performed: the accessible surface area was obtained by geometric method [15] using $N_2$ as a probe molecule of size 3.681 Å. This method consists of the probe molecule rolling over the framework. The pore volume of this structure was then estimated using the thermodynamic method developed by Myers and Monson [16], which considers a helium probe molecule (module as a LJ fluid $\sigma = 2.58$ Å ; $\epsilon/K_B = 10.22K$). The pore diameter was computed using the Geld and Gubbins method [1].

**DISCUSSION**

The Schwarzite structures considered in this work are presented in Figure 1. They have different pore sizes and shapes. In Table 1 we present the pore diameter of each structure considered here. IPWg presents the largest pore with a diameter of $d=14$ Å, while P688 has an ultra-small pore diameter of just $d=3.1$ Å. D688 and G688 have $d<3$ Å and can be considered as an almost non-porous material. The accessible surface area and the Helium pore volume are also displayed in Table 1. P7par has the largest surface area of 1529.05 $m^2/g$, followed by D8bal and G8bal 1263.88 and 1218.77 $m^2/g$, respectively. The P7par is also the one with the highest pore volume of 0.94 $cm^3/g$, followed by IWPg 0.9140 $cm^3/g$. All structures are considerably lightweight materials with mass density varying from 0.98 to 2.15 $g/cm^3$.

Table 1: Accessible surface area computed using the geometric method with $N_2$ probe molecule (size equal 3.681 Å) and Helium pore volume. The mass density values and the pore diameter are also presented.

|  | P688 | P7par | P8bal | D688 | D766 | D8bal | G688 | G8bal | IWPg |
|---|---|---|---|---|---|---|---|---|---|
| **Surface Area ($m^2/g$)** | - | 1529.05 | 1198.18 | - | 1068.47 | 1263.88 | - | 1218.77 | 1115.99 |
| **Helium Pore Volume ($cm^3/g$)** | 0.1588 | 0.9429 | 0.8561 | 0.0043 | 0.7131 | 0.8609 | 0.0 | 0.8295 | 0.9140 |
| **Pore diameter (Å)** | 3.1 | 8.3 | 3.6 / 9.2 | 2.1 | 5.2 / 6.6 | 4.7 / 6.4 | 0.8 | 4.6 | 4.1 / 14.3 |
| **Mass density ($g/cm^3$)** | 1.99 | 0.98 | 1.12 | 2.06 | 1.07 | 1.15 | 2.15 | 1.19 | 1.03 |

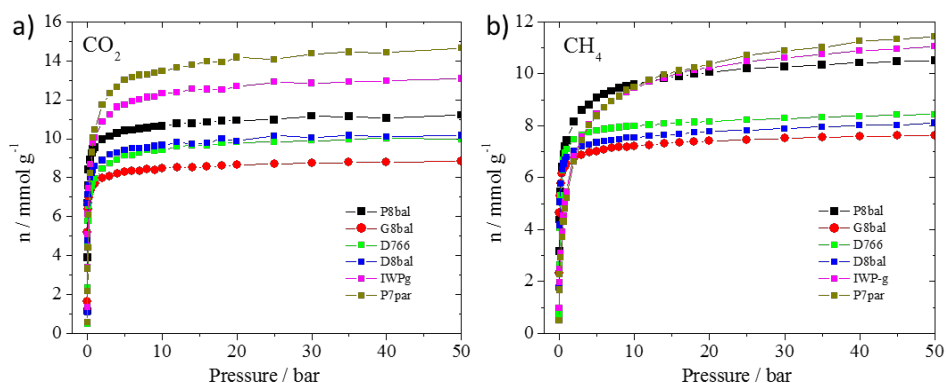

Figure 2: (a) CO2 and; (b) CH4 adsorption isotherms at T=303K. Curve ordering at 50 values (from bottom to up): (a) G8bal, D766, D8bal, P8bal, IWPg and P7par; (b) G8bal, D8bal, D766, P8bal, IWPg and P7par.

The gas sorption behavior was further investigated for each structure. GCMC simulations were carried out by varying the gas reservoir pressure to predict the number of molecules adsorbed in each solid at ambient temperature T=300K. In Figure 2 we present the single $CO_2$ (Figure 2a) and $CH_4$ (Figure 2b) component adsorption isotherms at 303 K for each studied case. All curves show a Type –I isotherm shape consistent with the behavior of a microporous adsorbent. Interestingly, the $CO_2$ uptake remains significantly higher than for $CH_4$ and other gas molecules for all explored solids. The higher $CO_2$ affinity is confirmed by the simulated adsorption enthalpies and it follows the same sequence as the gas uptake: $CO_2 > CH_4 > CO - N_2 > H_2$. The P7par is the best material with predicted adsorption uptakes of $CO_2$ at 303 K and 1 bar (~9.5 mmol/g) and 10 bar (~11.20 mmol/g), it surpasses the performances of other ultra-microporous materials previously envisaged for $CO_2$ capture [17]. This material is hydrophobic with water starting to be adsorbed only at pressure p/p0 > 0.3. The P7par is also the best material for $CH_4$ storage, with $CH_4$ storage capacity of 11.7 mmol/g. IPWg and P8bal also present very high $CH_4$ storage capacity (*i. e.,* 11.3 and 10.6 mmol/g, respectively). The high capacity of these materials is related to their high surface area and pore volume. From the energetic point of view, the IPWg and primitive family present the highest CO2 adsorption heat ~42 kJ/mol, while the gyroid and diamond families present adsorption heat ~39 kJ/mol.

As expected, the D688 and G688 do not adsorb any molecules and P688 is accessible only for $H_2$ molecules. The computed adsorption enthalpy of $H_2$ in P688 is 19.3 kJ/mol, which is much larger than all other materials studied here (~ 9 kJ/mol). Moreover, this value is exceptional high when compared with metal-organic frameworks (5-9 kJ/mol) [17,18] and other carbon surfaces such as carbon foams and nanotubes (3-7 kJ/mol) [19]. Further GCMC simulations at different temperature were performed. At temperature 200 K, P688 achieves its full capacity at pressure < 10bar (see Figure 3b). Due the ultra-small pore size P688 can store only one single $H_2$ molecule per cage (see snapshot of Figure 3a). Although the $H_2$ storage capacity in P688 is relatively low (~3.4 mmol/g), it is important to remark that P688 is saturated at relatively high temperature and low pressure (*i.e.*, 200K; 10bar). Other porous materials present $H_2$ storage capacity up to 10 times larger than P688, but it is possible only at 77K and pressure > 100 bar [17]. Further analysis on the arrangement and the explanation of the strong $H_2$-P688 interaction energy will be provided in our future publication.

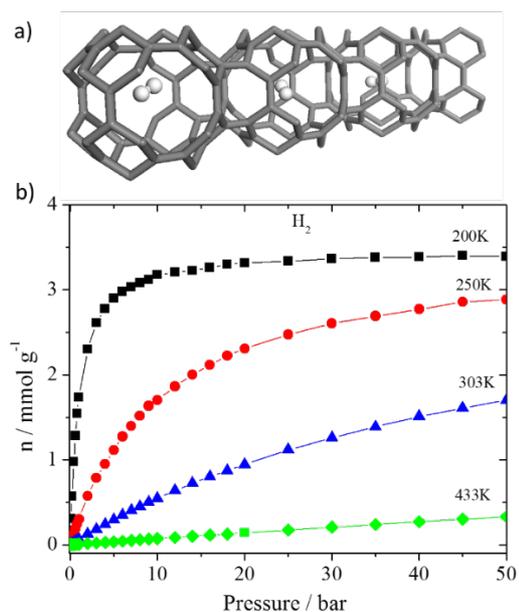

Figure 3: (a) MD snapshopt of adorbed H2 into P688. (b) H2 adsorption isotherm in in P688 at different temperature. The adsorption enthalpy is 19kJ/mol

**CONCLUSION**

Computational studies were performed to study the nature gas adsorption properties in a series of carbon porous materials of Schwarzite families. Gas adsorption isotherms and enthalpies were predicted. Our systematic study shows promising features for Schwarzites in gas storage applications, some of the structures presenting exceptional performances. The best candidates are P7par, P8bal and IWPg that have considerably high $CO_2$ and $CH_4$ storage capacity. P688 presents significant high $H_2$ heat adsorption that make this material interesting for H2 storage at relatively high temperature. Further validation of these results, as well as the molecular understanding of adsorbed-adsorbent interactions will be provided in our future publication [20].

**ACKNOWLEDGEMENTS**

The authors also thank the Center for Computational Engineering and Sciences at Unicamp for financial support through the FAPESP/CEPID Grant # 2013/08293-7. DDB thanks FAPESP Grant # 2015/14703-9 for financial support.